\documentclass[aps,nofootinbib,a4paper,11pt]{revtex4-2}
\usepackage{amsfonts,amsmath,amssymb}
\usepackage{graphicx}
\usepackage{hyperref}
\usepackage{color}
\usepackage{slashed}
\usepackage[usenames,dvipsnames]{xcolor}
\def\trmin{{\rm tr}}
\DeclareMathOperator\arctanh{arctanh}
\DeclareMathOperator\arccoth{arccoth}
\DeclareMathOperator\arcsinh{arcsinh}
\DeclareMathOperator*{\sumint}{%
\mathchoice%
  {\ooalign{$\displaystyle\sum$\cr\hidewidth$\displaystyle\int$\hidewidth\cr}}
  {\ooalign{\raisebox{.14\height}{\scalebox{.7}{$\textstyle\sum$}}\cr\hidewidth$\textstyle\int$\hidewidth\cr}}
  {\ooalign{\raisebox{.2\height}{\scalebox{.6}{$\scriptstyle\sum$}}\cr$\scriptstyle\int$\cr}}
  {\ooalign{\raisebox{.2\height}{\scalebox{.6}{$\scriptstyle\sum$}}\cr$\scriptstyle\int$\cr}}
}

\begin{document}

\title{\Large{Light pseudo-scalar meson masses under strong magnetic fields within the SU(3) Nambu-Jona-Lasinio model}}

\author{Sidney S. Avancini\,} \email{sidney.avancini@ufsc.br}
\author{Joana C. Sodr\'e\,} \email{joana.sodre@gmail.com}
\affiliation{Departamento de F\'{\i}sica, Universidade Federal de Santa Catarina, 88040-900 Florian\'{o}polis, Santa Catarina, Brazil}
\author{M\'aximo Coppola\,}\email{coppola@tandar.cnea.gov.ar}
\author{Norberto N. Scoccola\,}\email{scoccola@tandar.cnea.gov.ar}
\affiliation{CONICET, Rivadavia 1917, (1033) Buenos Aires, Argentina}
\affiliation{Physics Department, Comisi\'on Nacional de Energ\'ia At\'omica,
Av. Libertador 8250, (1429) Buenos Aires, Argentina \vspace*{1cm}}

\begin{abstract}
{\centering \large Abstract \par}

We calculate the pole masses of pseudoscalar
mesons in a strongly magnetized medium within the framework of the
SU(3) Nambu-Jona--Lasinio model, using a magnetic
field-independent regularization scheme. We employ both a constant and a magnetic
field-dependent coupling $G(B)$,  the latter being
fitted to reproduce lattice QCD results for the pseudocritical
chiral transition temperature. Numerical results for the pole
masses are obtained for definite parametrizations of the model.
For neutral mesons, the use of $G(B)$ provides closer agreement
with lattice QCD results, which reveal a decrease of the mass with
the external field. On the contrary, charged mesons masses are
enhanced by $B$, showing no sign of the non-monotonous behavior
found in recent lattice QCD simulations. 
\end{abstract}

\maketitle
\clearpage

\section{Introduction}

The behavior of strongly interacting matter under the influence of intense magnetic
fields has been attracting a lot of interest in recent years.
This interest is partly motivated by the fact that
strong magnetic fields have been achieved or hypothesized in
several physical situations. For example, in non-central
relativistic heavy-ion collisions, magnetic fields of magnitude as
large as $B\sim 10^{19}-10^{20}$ G are generated due
to the motion of charged spectator particles, essentially at the
earliest times of the collision~\cite{Tuchin:2013ie}. In addition,
strong magnetic fields may also play an important role in
astrophysics scenarios, such as matter formation in the early
Universe~\cite{Grasso:2000wj} or in the dynamics of magnetars,
where the inner core can possibly harbor magnetic fields strengths
as large as $B \sim 10^{19}$ G~\cite{Duncan:1992hi}. One expects
that new and detectable effects in the phase diagram and
properties of strongly interacting matter will emerge due to these
extreme magnetic fields, causing numerous phenomenological
consequences. For example, the Chiral Magnetic
Effect~\cite{Fukushima:2012vr,Kharzeev2014,Li:2020dwr},
Chiral Separation Effect~\cite{Gorbar:2013upa}, Chiral Magnetic
Wave~\cite{Kharzeev:2010gd,Shovkovy:2018tks,Burnier:2011bf,Yee:2013cya,STAR:2015wza}
and related phenomena are supposed to be experimental signals of
the influence of strong magnetic fields in the QCD
matter~\cite{Fukushima:2008xe,Kharzeev:2015znc}.

% IMC
From the theoretical point of view, first principle analytical QCD
calculations are very difficult to perform in the
non-perturbative regime given the complexity of the theory.
Therefore, one has to make use of alternative procedures to tackle
the problem. In this regard, great progress has been made in
recent years on the investigation of the QCD phase diagram by
using either lattice QCD (LQCD) simulations or effective models,
which can work together in a complementary manner in the face of
lack of experimental evidence in some observables. In fact, in
many of these models available experimental or LQCD results are
used to fix their phenomenological parameters, allowing for
improved results. In particular, results from LQCD calculations at
zero temperature and physical pion masses show that an external
constant magnetic field enforces the quark condensate favoring the
breakdown of chiral symmetry, an effect known as magnetic
catalysis (MC)~\cite{DElia:2010abb}. This result is in agreement
with most effective model
calculations~\cite{Andersen:2014xxa,Miransky:2015ava}. At finite
temperature, the majority of these models predict the increase of
the pseudo-critical transition temperature $T_{pc}$ with the magnetic
field~\cite{Andersen:2014xxa,Miransky:2015ava}. Nevertheless,
accurate LCQD
results~\cite{Bali:2011qj,Bali:2012zg,Bali2014,Bornyakov2014} have
shown the opposite pattern; $T_{pc}$ decreases with the magnetic
field, a phenomenon dubbed as inverse magnetic catalysis (IMC).
The explanation for IMC at finite temperature is still
controversial and under study [see
Ref.~\cite{Bandyopadhyay:2020zte} for a mini-review on the IMC
effect]. It is not clear that the nonmonotic behavior of the quark
condensate is in fact the mechanism behind the IMC
effect~\cite{DElia:2018xwo,Endrodi:2019zrl,Andersen:2021lnk}. In
the context of effective models, several possibilities have been
explored in the recent literature to incorporate the IMC effect
phenomenologically. Within the Nambu-Jona--Lasinio (NJL) model for
example, which we will use in this work, these improvements
include going beyond mean-field calculations~\cite{Mao:2016fha} or
taking into consideration the anomalous magnetic moment of
quarks~\cite{Fayazbakhsh:2014mca,Mei:2020jzn,Chaudhuri:2020lga,Xu:2020yag}.
Motivated by the running of the QCD coupling, one of the simplest
modifications available consists of introducing a coupling
constant that  depends on the magnetic field (and in some cases also on the temperature) and can be fixed by fitting some LQCD results,
such as the quark condensate or the chiral pseudocritical temperature. This strategy has shown that the NJL model can satisfactorily
reproduce LQCD results in a broad range of
temperature and magnetic
fields~\cite{Ferreira:2014kpa,Farias:2014eca,Farias:2016gmy,Avancini:2016fgq,Avancini:2018svs,Avancini:2020xqe}.
In this regard, an interesting possibility was recently proposed
in Ref.~\cite{Endrodi:2019whh}. There,  the magnetic field dependent four fermion
coupling is fitted to reproduce constituent quark masses, which
are obtained from the LQCD calculation of baryon masses by
assuming in a simplified way that the baryon mass can be obtained
by merely summing the masses of their constituents. Lastly,
calculations using the non-local NJL model have shown that IMC
is obtained naturally~\cite{Pagura2017,Dumm2017}.

% Meson masses
The presence of strong magnetic fields also has a meaningful
impact on hadron properties. In this work we will focus on its
consequences over the masses of the light pseudoscalar meson
nonet, which has drawn a lot of attention in recent years. Most
calculations in the literature have been performed for the
lightest mesons. 
For pions at zero temperature, LQCD simulations show an overall decrease of the neutral pion mass with the magnetic field (both in quenched QCD and using  staggered fermions), while charged pions exhibit the opposite behavior within the quenched approximation~\cite{Luschevskaya:2015cko,Bali:2017ian}. A recent simulation using highly improved staggered fermions with a slightly heavier-than-physical pion mass of 220 MeV was performed in Ref.~\cite{Ding2021}, where the masses of many pseudoscalar mesons are computed. There, the decreasing trend of the neutral pion (and kaon) mass is confirmed. Moreover, charged pions (and kaons) reveal an initial increase with the magnetic field up to values $eB\sim 0.6$ GeV$^2$, in accordance with previous results from Ref.~\cite{Bali:2011qj} where $eB<0.5$ GeV$^2$ values are considered for the charged pion mass using stout smeared staggered fermions. However, in stark contrast with previous quenched results from Refs.~\cite{Luschevskaya:2015cko,Bali:2017ian}, for stronger magnetic fields this increasing tendency is found to be reversed, resulting in a nonmonotonous behavior.

On the other hand, the influence of magnetic fields on the
lightest scalar and pseudoscalar mesons ($\sigma$ and $\pi$) has
also been calculated mostly using two-flavor schemes, such as
chiral perturbation
theory~\cite{Agasian:2001ym,Andersen:2012zc,Orlovsky:2013gha,Colucci:2013zoa},
the linear sigma model~\cite{Ayala:2018zat,Das2020,Ayala:2020dxs},
two-flavor quark-meson model~\cite{Kamikado:2013pya}, relativistic
hamiltonian based
formalisms~\cite{Orlovsky:2013gha,Andreichikov:2016ayj}, effective
chiral confinement Lagrangian
approach~\cite{Simonov:2015xta,Andreichikov:2018wrc}, QCD sum
rules~\cite{Dominguez:2018njv}, the two-flavor NJL
model~\cite{Fayazbakhsh:2012vr,Fayazbakhsh:2013cha,Avancini:2015ady,Avancini:2016fgq,Coppola:2018vkw,Coppola:2019uyr,Zhang:2016qrl,Mao:2017wmq,Mao:2018dqe,Wang:2017vtn,Liu:2018zag}
or its non-local version~\cite{GomezDumm:2017jij,Dumm:2020muy}. In
this context, there are very few calculations of meson properties
incorporating the strange quark. In Refs.~\cite{Hattori:2015aki,Kojo:2021gvm}, using a
non-relativistic constituent SU(3) quark model, neutral and
charged mesons masses are considered. 
By using a relativistic hamiltonian based formalism,
in Refs.~\cite{Orlovsky:2013gha,Andreichikov:2016ayj} pions and
kaons are calculated and comparisons with chiral perturbation
theory and LQCD results are considered. In
Ref.~\cite{Mishra:2020ckq}, kaons and anti-kaons are investigated
in a chiral SU(3) model.

% NJL model
In order to study the behavior of the masses of the light
pseudoscalar meson nonet in the presence of an external constant
magnetic field, we will use the SU(3) NJL model. We remind that
the NJL model is a non-renormalizable model and a regularization
procedure has to be adopted, which may be considered as part of
the definition of the effective model. In fact, the choice of an
appropriate regularization scheme is a crucial issue for the
description of physical systems. It has been shown that the use of
an inappropriate regularization scheme causes strong oscillations
in meson masses and tachyonic or discontinuous behavior of masses.
When working with quark matter immersed in a magnetized medium,
performing an exact separation of magnetic from non-magnetic
contributions for all physical observables is a key point for
their correct description, a strategy known as the ``magnetic
field independent regularization'' (MFIR) scheme. The importance
of the regularization procedure has been reviewed in
Ref.~\cite{Avancini:2019wed}, where it is shown that the MFIR
scheme is free of these unphysical behaviors, which are due to an
improper regularization. An improvement within the MFIR scheme was
recently suggested in Ref.~\cite{Avancini:2020xqe} for the
calculation of many mean-field observables. However, this modification is not relevant for the quantities we will study in this
work, namely quark condensates and meson masses, so we can safely
omit it.

Regarding the determination of meson masses within the NJL model,
one important point is the proper calculation of charged mesons.
In this case, polarization functions have to be carefully handled
in order to be diagonalized, since Schwinger phases arising from
quarks propagators do not cancel, leading to a breakdown of
translational invariance. As shown in Ref.~\cite{Coppola:2018vkw},
an appropriate treatment involves the use of the Ritus
basis~\cite{Ritus:1978cj}. Unfortunately, this issue has not
been properly addressed in several calculations. Of course, for
neutral mesons the usual momentum basis can be used since the
Schwinger phase factor cancels out in that case. In this regard,
the NJL model shows an enhancement of the neutral pion mass for
sufficiently strong magnetic fields, in contradiction with lattice
results. One possible approach to overcome this issue, which we
will adopt in this work, is to introduce a  magnetic field dependent
coupling constant, determined by fitting LQCD results for the quark
condensate, which, as mentioned earlier, also allows the model to
incorporate the IMC effect at finite temperatures. Results
obtained following this strategy agree very well with LQCD
simulations~\cite{Avancini:2015ady,Avancini:2016fgq}. An
alternative approach comprises the use of the non-local version of
the SU(2) NJL model, where the neutral pions mass has been shown
to naturally decrease with the magnetic field in agreement with
LQCD results~\cite{GomezDumm:2017jij,Dumm:2020muy}.

As mentioned earlier, there are very few calculations of meson
properties incorporating the strange quark. The aim of the present
paper is to study the behavior of the pseudoscalar meson nonet
masses as functions of the magnetic field. To that end we use the
SU(3) NJL model including the 't Hooft-Maekawa interaction which
breaks the U$_{\rm A}$(1) symmetry. We work within the MFIR scheme
and consider both the case of a fixed and of a $B$-dependent
four-fermion coupling constant. At the mean-field level we
calculate quark condensates, which are compared with LCQD results
to find qualitative agreement. For the calculation of the light
pseudoscalar meson nonet we adopt the RPA formulation, where
special care has to be taken to the fact that constituent quark
masses are different for each flavor. For charged mesons the Ritus
basis is used to diagonalize the polarizers, resulting in
monotonically increasing masses for both constant and magnetic
couplings. For neutral mesons, the
polarizers calculation is simplified since Schwinger phases
cancels out. Note that the 't Hooft-Maekawa
interaction together with the uniform magnetic field $B$ induce a
mix of neutral mesons states with equal flavors, i.e.
$\pi^0,\eta,\eta'$ \cite{Cao2021}. We see that, except for
$\eta^\prime$, neutral pseudoscalar mesons display a
non-monotonous behavior when using a constant coupling, which
shifts to a monotonous decrease in concordance with LCQD results
when a magnetic coupling is introduced. As already known from the
usual SU(3) NJL model at $B=0$, the $\eta^\prime$ meson comes out
in the model as a resonance or unstable particle. In this case,
the propagator becomes a complex number and from the analysis of
the complex pole, the mass of the resonance
is obtained. This situation gets intricate when the magnetic field is present, thus, we have
developed a new formalism to treat this case.

We organize this work as follows. In Sec.~II we introduce the
theoretical formalism used to obtain neutral and charged
pseudoscalar meson masses. Then, in Sec.~III we present and
discuss our numerical results, while in Sec.~IV we provide a
summary of our work, together with our main conclusions. We also
include Appendices~A and~B to quote some technical details of our
calculations.

\section{Theoretical formalism}

\subsection{Effective Lagrangian and mean field properties}

We consider the Euclidean action of the SU(3) NJL model which includes a
scalar-pseudoscalar interaction and the 't Hooft six-fermion
interaction in the presence of an external magnetic field. It is written as
\begin{equation}\label{lf}
S_E = \int d^4x  \left[ \bar{\psi}\left( -i\ \rlap/\!D  + \hat{m} \right)\psi - G\sum_{a=0}^{8}\left[ \left( \bar{\psi}\lambda_{a}\psi \right)^{2}
    +\left( \bar{\psi}i\gamma_{5}\lambda_{a}\psi \right)^{2} \right]
+K \left( d_{+}+d_{-}\right)\right]\, ,
\end{equation}
\noindent where $G$ and $K$ are coupling constants, $\psi
=\left(\psi_u,\psi_d,\psi_s\right)^{T}$ represents a quark field with three
flavors, $d_{\pm}=\det\left[ \bar{\psi}\left(1\pm\gamma_{5} \right)\psi \right]$ and $\hat{m}=\mathrm{diag}\left( m_{u},m_{d},m_{s} \right)$ is the
corresponding current quark mass matrix. In addition,  $\lambda_{0}=\sqrt{2/3}\,I$,
where $I$ is the unit matrix in the three flavor space, and
$\lambda_{a}$ with $a=1,...,8$ denote the Gell-Mann matrices. The
coupling of quarks to the electromagnetic field ${\cal A}_\mu$
is implemented through the covariant derivative
$D_{\mu}=\partial_\mu - i \hat Q {\cal A}_{\mu}$ where $\hat Q=\mathrm{diag}\left( Q_{u},Q_{d},Q_{s} \right)$ represents the quark electric charge matrix with $Q_u/2 = -Q_d = - Q_s = e/3$, $e$ being the proton electric charge. In the present work we consider a static and
constant magnetic field in the $3$-direction. Using the Landau gauge we have ${\cal
A}_\mu=\delta_{\mu 2} x_1 B$.

In order to study meson properties, we proceed by bosonizing the
action in terms of scalar  $\sigma_a(x)$ and pseudoscalar
$\pi_a(x)$ fields and the corresponding auxiliary
$\mbox{s}_a(x)$ and $\mbox{p}_a(x)$ fields.
Following the standard procedure, we start with the partition
function
\begin{equation}
 Z=\int D\bar{\psi} D\psi \ e^{-S_E} \, .
\end{equation}
By introducing functional delta functions, the scalar
($\bar{\psi}\lambda_{a}\psi$) and pseudoscalar
($\bar{\psi}i\gamma_{5}\lambda_{a}\psi$)  terms  present in $S_E$
are  replaced by $\mbox{s}_a(x)$ and $\mbox{p}_a(x)$
and the functional integration on the fermionic fields $\psi$ and
$\bar{\psi}$  can be performed by standard methods.
%Next, within the stationary phase approximation (SPA) the bosonized action can be conveniently written in terms of the bosonic fields.
To perform the integration over the auxiliary fields we use the
stationary phase approximation (SPA), choosing 
$\tilde{\mbox{s}}_a(x)$ and $\tilde{\mbox{p}}_a(x)$ in order to
minimize the integrand of the partition function. This yields a
set of coupled equations among the bosonic fields; at the end,
 $\tilde{\mbox{s}}_a(x)$ and $\tilde{\mbox{p}}_a(x)$
are to be considered as implicit functions of
$\sigma_a(x)$ and $\pi_a(x)$. Finally, we use the
mean field approximation by expanding the bosonized action in
powers of field fluctuations around the corresponding
translationally invariant mean field values $\bar{\sigma}_a$ and
$\bar{\pi}_a$, i. e.,
$\sigma_a(x)=\bar{\sigma}_a+\delta\sigma_a(x)$ and
 $\pi_a(x)=\bar{\pi}_a+\delta\pi_a(x)$. Due to charge conservation, only
 $\bar{\sigma}_0$, $\bar{\sigma}_3$ and $\bar{\sigma}_8$ are different from zero,
 while the vacuum expectation values of pseudoscalar boson fields are zero, $\bar{\pi}_a=0$.
 For convenience, we introduce $\bar{\sigma}=\mathrm{diag}(\bar{\sigma}_u,\bar{\sigma}_d,\bar{\sigma}_s)=\lambda_0 \bar{\sigma}_0 + \lambda_3\bar{\sigma}_3+\lambda_8\bar{\sigma}_8$.
At the mean field level, the Euclidean action per unit volume reads
\begin{equation}
\dfrac{\bar{S}_{\! E}^{\; \mathrm{bos}}}{V^{(4)}} \ =  -
\dfrac{N_c}{V^{(4)}} \sum_{f=u,d,s} \int d^4x \, d^4x' \
\trmin_D\,\ln \left(\mathcal{S}^f_{x,x'}\right)^{-1} -\dfrac{1}{2}
\left[ \bar{\sigma}_f \ \bar{\mbox{s}}_f + G\ \bar{\mbox{s}}_f \
\bar{\mbox{s}}_f - \dfrac{K}{2}\, \bar{\mbox{s}}_u \
\bar{\mbox{s}}_d \ \bar{\mbox{s}}_s \right] \, ,
%+2G\left( \phi_{u}^{2}+\phi_{d}^{2}+\phi_{s}^{2} \right)-4K\phi_{u}\phi_{d}\phi_{s} \, ,
%\label{seff}
\end{equation}
where $\trmin_D$ stands for the trace in Dirac space while
$\left(\mathcal{S}^f_{x,x'}\right)^{-1} = \delta(x-x')\left[ -i (
\slashed \partial - i Q_f \slashed {\cal A} ) + M_f \right]$
represents the inverse mean field quark propagator for each flavor
with effective mass $M_f=m_f+\bar{\sigma}_f$. 
Moreover, $\bar{\mbox{s}}_f = \tilde{\mbox{s}}_f(\bar \sigma_a)$
represent the auxiliary fields at the mean field level within the
SPA approximation (note that $\bar{\mbox{p}}_f=0$). From
the condition $ \delta \bar{S}_{\! E}^{\; \mathrm{bos}} / \delta
\bar{\sigma}_f =0$ it follows that $\bar{\mbox{s}}_f =  2\phi_f$
where $\phi_f$ is the chiral condensate for each flavor given by
\begin{equation}
 \phi_f= \langle{\bar \psi}_{f} \psi_{f}\rangle = -\dfrac{\delta \bar{S}^{\; bos}_{\! E}}{\delta m_f} = - \dfrac{N_c}{V^{(4)}} \int d^4x \ \trmin_D \ \mathcal{S}^f_{x,x} \ ,
\end{equation}
As is well known, the quark propagator can be written in different ways~\cite{Andersen:2014xxa,Miransky:2015ava}.
For convenience we take the following one
\begin{equation}
\mathcal{S}^f_{x,x'} \ = \ e^{i\Phi_f(x,x')}\,\int_p e^{i p\, (x-x')}\, \tilde{\mathcal{S}}_p^f \, ,
\label{sfx}
\end{equation}
where $\Phi_f(x,x')= Q_f B (x_1+x_1')(x_2-x_2')/2$ is the
so-called Schwinger phase. We have introduced here the shorthand notation
\begin{equation}
\int_{p}\ \equiv \ \int \dfrac{d^4 p}{(2\pi)^4}\ .
\label{notation1}
\end{equation}
We express $\tilde{\mathcal{S}}_p^f$ in the Schwinger form~\cite{Andersen:2014xxa,Miransky:2015ava}
\begin{align}
\tilde{\mathcal{S}}_p^f \, =& \int_0^\infty \!d\tau\,
\exp\left[-\tau \left( M_f^2 + p_\parallel^2 +
\dfrac{\tanh(\tau B_f)}{\tau B_f}\; p_\perp^2\ - i \epsilon \right) \right] \times \nonumber\\[2mm]
& \left\{\left(M_f-p_\parallel \cdot \gamma_\parallel \right)
\, \left[1+i s_f \,\gamma_1 \gamma_2\, \tanh(\tau B_f)\right] -
\dfrac{p_\perp \cdot \gamma_\perp}{\cosh^2(\tau B_f)} \right\} \ ,
\label{sfp_schw}
\end{align}
where the following definitions have been used. The ``perpendicular'' and
``parallel'' gamma matrices are collected in vectors $\gamma_\perp =
(\gamma_1,\gamma_2)$ and $\gamma_\parallel = (\gamma_3,\gamma_4)$.
Similarly, $p_\perp = (p_1,p_2)$ and $p_\parallel = (p_3,p_4)$. Note that in
our convention $\{\gamma_\mu,\gamma_\nu\}=-2 \delta_{\mu\nu}$  and we have introduced the
notation $s_f = {\rm sign} (Q_f B)$ and $B_f=|Q_fB|$. The limit $\epsilon\rightarrow 0$ is implicitly understood.

The integral in Eq.~\eqref{sfp_schw} is divergent and has to be
properly regularized. We will use the MFIR scheme, where one subtracts from the unregulated integral the
$B = 0$ limit and then adds it in a regulated form. We obtain
\begin{equation}
\phi_f^{reg} = \phi_f^{vac\vphantom{g}} + \phi_f^{mag} \, , \qquad
\begin{cases}
\phi_f^{vac\vphantom{g}} \equiv -N_c M_f \, I_{1f}^{vac\vphantom{g}} \\
\phi_f^{mag} \equiv -N_c M_f \,I_{1f}^{mag}
\end{cases} \, .
\label{phif}
\end{equation}
The expression of $I_{1f}^{vac\vphantom{g}}$ for the 3D cutoff regularization scheme we use in this work can be found in Eq.~\eqref{I1freg} of App.~A. The expression of $I_{1f}^{mag}$, given in Eq.~\eqref{I1magAppB} of App.~B, reads
\begin{equation}
I_{1f}^{mag} = \dfrac{B_f}{2\pi^2} \left[ \ln \Gamma(x_f) - \left(x_f - \dfrac{1}{2}\right) \ln x_f + x_f - \dfrac{\ln{2\pi}}{2} \right] \, , \label{i1}
\end{equation}
where $x_f=M_f^2/(2B_f)$.

Finally, by combining the equations from the SPA together with the
gap equations, we obtain that the regularized form of the  set
of coupled equations for the effective quarks masses read
\begin{align}
 M_{u} &= m_u - 4 G\ \phi_u^{reg} + 2K\ \phi_d^{reg}\phi_s^{reg} \ , \nonumber\\
 M_{d} &= m_d - 4 G\ \phi_d^{reg} + 2K\ \phi_s^{reg}\phi_u^{reg} \ , \nonumber\\
 M_{s} &= m_s - 4 G\ \phi_s^{reg} + 2K\ \phi_u^{reg}\phi_d^{reg} \ .
\label{gapeqs}
\end{align}

\subsection{Meson sector}

For the calculation of meson masses, we consider the  second-order correction to the mean field bosonized Euclidean action ${S}_E$.
At the quadratic level we get for the pseudoscalar sector
\begin{equation}
S^{\,\mbox{\tiny quad}}_{mes} \ = \ \dfrac{1}{2}  \int d^4 x' d^4x
\sum_{P,P'} \ \delta P^*(x)\ {\cal G}_{P,P'}(x,x')\ \delta P'(x')
\, ,
\end{equation}
where the sum indexes run over the nonet of pseudoscalar mesons.
Namely, $P,P' = \pi_3, \pi^\pm, K^0, \bar K^0,$ $K^\pm, \eta_0, \eta_8$.
The inverse meson propagator in coordinate space can be written as
\begin{equation}
{\cal G}_{P,P'}(x,x') = T_{P,P'} \ \delta^{(4)}(x-x') -
J_{P,P'}(x,x') \, .
\end{equation}

For $P, P'= \pi^\pm, K^\pm, K^0, \bar K^0$ this operator is diagonal
\begin{equation}
T_{P,P'} = T_P \ \delta_{P,P'} \qquad , \qquad J_{P,P'}(x,x') =
J_P(x,x') \ \delta_{P,P'}\, ,
\end{equation}
where
\begin{alignat}{6}
T_{\pi^+} &= T_{\pi^-} &= \left[ 2 G - K \phi_s \right]^{-1}   \qquad  &, \qquad J_{\pi^+}(x,x') &= J_{\pi^-}(x',x) &= c_{ud}(x,x')\, , \\
T_{K^+}   &= T_{K^-}   &= \left[ 2 G - K \phi_d \right]^{-1}   \qquad  &, \qquad  J_{K^+}(x,x') &= J_{K^-}(x',x) &= c_{us}(x,x')\, , \\
T_{K^0} &= T_{\bar K^0} &= \left[ 2 G - K \phi_u \right]^{-1}   \qquad  &, \qquad   J_{K^0}(x,x') &= J_{\bar K^0}(x',x) &= c_{ds}(x,x')\, .
\end{alignat}
In these expressions
\begin{equation}
c_{ff'}(x,x') = 2 N_c \ tr_D \left[ \mathcal{S}^f_{x,x'} \ \gamma_5 \  \mathcal{S}^{f'}_{x',x} \ \gamma_5 \right] \, .
\label{cff'}
\end{equation}

On the other hand, the two-point function ${\cal G}_{P,P'}(x,x')$
is non-diagonal but symmetric in the $P, P' = \pi_3, \eta_0,
\eta_8$ subspace. The corresponding matrix elements of $T_{P,P'}$
are
\begin{eqnarray}
T_{\pi_3\pi_3} &=& \dfrac{ K^2 \left( \phi_u + \phi_d \right)^2 - 4 G K \phi_s - 8 G^2}{f}\, , \nonumber\\[2mm]
T_{\eta_0\pi_3}&=& \dfrac{ 2 \left[ K^2 ( \phi_u + \phi_d - \phi_s) - 2 G K \right]
\left( \phi_u - \phi_d \right) }{\sqrt{6} f}\, ,  \nonumber\\[2mm]
T_{\eta_8\pi_3}&=& \dfrac{ \left[ K^2  ( \phi_u + \phi_d + 2 \phi_s) + 4 G K \right]
\left( \phi_u - \phi_d \right) }{\sqrt{3} f}\, ,  \nonumber\\[2mm]
T_{\eta_0\eta_0} &=& \dfrac{ 2 K^2 \left[ ( \phi_d - \phi_s)^2 + \phi_u ( \phi_u - 2\phi_d - 2\phi_s) \right]
+ 8 G K ( \phi_u + \phi_d + \phi_s) - 24 G^2 }{3 f}\, , \nonumber \\[2mm]
T_{\eta_8\eta_0}&=&  \dfrac{ 2 K^2 \left[ ( \phi_u - \phi_d )^2 + \phi_s ( \phi_u + \phi_d - 2 \phi_s )\right]
- 4 G K \left( \phi_u + \phi_d - 2 \phi_s \right) }{3 \sqrt{2} f}\, ,  \nonumber\\[2mm]
T_{\eta_8\eta_8} &=& \dfrac{ K^2 \left[ ( \phi_u - \phi_d )^2 + 4 \phi_s ( \phi_u + \phi_d + \phi_s) \right]
- 4 G K \left( 2 \phi_u + 2 \phi_d - \phi_s \right) - 24 G^2 }{ 3 f} \, ,
\label{eq8}
\end{eqnarray}
where
\begin{equation}
f= -4 K^3 \phi_u \phi_d \phi_s + 4 G K^2 \left( \phi_u^2 + \phi_d^2 + \phi_s^2 \right) - 16 G^3 \, .
\end{equation}
In turn, the polarization function elements can be expressed as
\begin{equation}
J_{P,P'}(x,x') = \sum_f \gamma^f_{P,P'} \ c_{ff}(x,x') \, ,
\end{equation}
where the coefficients $\gamma^f_{P,P'}$ are given by
\begin{alignat}{8}
\gamma^u_{\pi_3\pi_3} &= +\gamma^d_{\pi_3\pi_3} &&= \dfrac{1}{2} \quad &,
\qquad \gamma^s_{\pi_3\pi_3} &= 0 \quad &,
\qquad \gamma^u_{\eta_0\eta_0} &= \gamma^d_{\eta_0\eta_0} = \gamma^s_{\eta_0\eta_0} &&= \dfrac{1}{3} \ ,
\nonumber \\[3mm]
\gamma^u_{\eta_0\pi_3} &= -\gamma^d_{\eta_0\pi_3} &&= \dfrac{1}{\sqrt 6} \quad &,
\qquad \gamma^s_{\eta_0\pi_3} &= 0 \quad &,
\qquad \gamma^u_{\eta_8\eta_0} &= \gamma^d_{\eta_8\eta_0} = -\dfrac{1}{2} \gamma^s_{\eta_8\eta_0} &&= \dfrac{1}{3\sqrt2} \ ,
\nonumber \\[3mm]
\gamma^u_{\eta_8\pi_3} &= -\gamma^d_{\eta_8\pi_3} &&= \dfrac{1}{2\sqrt3} \quad &,
\qquad \gamma^s_{\eta_8\pi_3} &= 0 \quad &,
\qquad \gamma^u_{\eta_8\eta_8} &= \gamma^d_{\eta_8\eta_8} = \dfrac{1}{4}\gamma^s_{\eta_8\eta_8} &&= \dfrac{1}{6} \ .
\label{gammas}
\end{alignat}

\subsubsection{Neutral mesons}

For neutral mesons the contributions of Schwinger phases associated with the quark
propagators in Eq.~\eqref{cff'} cancel out. Therefore, the polarization functions depend only on the difference
$(x - x')$, which leads to the conservation of momentum, since they are translationally invariant.
If we take the Fourier transform of neutral meson fields to the momentum basis, the corresponding
transform of the polarization functions will be diagonal in momentum space. Thus, the
neutral meson contribution to the quadratic action in the momentum basis can be written as
\begin{align}
S^{\,\mbox{\tiny quad}}_{neut.mes} \ =&
\ \dfrac{1}{2}  \int_q \sum_{P=K^0,\bar K^0} \ \delta P^*(-q)\ {\cal G}_{P}(q_\perp^2,q_\parallel^2)\ \delta P(q) \nonumber \\[2mm]
& + \ \dfrac{1}{2}  \int_q \sum_{P,P'=\pi_3,\eta_0,\eta_8} \
\delta P^*(-q)\ {\cal G}_{P,P'}(q_\perp^2,q_\parallel^2)\ \delta
P'(q) \, .
\end{align}
Here, the inverse neutral kaon propagator is given by
\begin{align}
{\cal G}_{K^0}(q_\perp^2,q_\parallel^2)={\cal G}_{\bar K^0}(q_\perp^2,q_\parallel^2)=
\left[ 2 G - K \phi_u \right]^{-1} - c_{ds}(q_\perp^2,q_\parallel^2) \, , \label{GK0}
\end{align}
while for  $P,P'=\pi_3,\eta_0,\eta_8$ we have
\begin{align}
{\cal G}_{P,P'}(q_\perp^2,q_\parallel^2) = T_{P,P'} + \sum_f
\gamma^f_{P,P'}\ c_{ff}(q_\perp^2,q_\parallel^2) \, . \label{Gff}
\end{align}
The values of  $T_{P,P'}$ and  $\gamma^f_{P,P'}$ can be found in
Eqs.~\eqref{eq8} and~\eqref{gammas}, respectively.

In the neutral case, the functions $c_{ff'}$ in momentum space are given by
\begin{align}
c_{ff'}(q_\perp^2,q_\parallel^2) =
2 N_c \int_p tr_D \left[ \tilde{\mathcal{S}}_{p_-}^f \ \gamma_5 \  \tilde{\mathcal{S}}_{p_+}^{f'} \ \gamma_5 \right]\, , \label{cff'mom}
\end{align}
where $p^\pm = p \pm q/2$.
%Details of the calculation $c_{ff'}(q_\perp^2,q_\parallel^2)$ can be found in App.B.
We remark here that these functions are divergent. Within the MFIR scheme they can be regularized as
\begin{align}
c^{reg}_{ff'}(q_\perp^2,q_\parallel^2) = c^{vac\vphantom{g}}_{ff'}(q^2) +  c^{mag}_{ff'}(q_\perp^2,q_\parallel^2)\, , \label{cff'neutro}
\end{align}
where the first term in the right-hand side correspond to the
vacuum contribution while the second term to the magnetic one. In
this work we regularize the otherwise divergent vacuum term
through a 3D cutoff; the corresponding expression is given in
Eq.~\eqref{cff'reg} of App.~A. For the calculation of the masses
we can set $q_\perp^2=0$, while the Euclidean parallel components
are to be evaluated at the negative real space
$q_\parallel^2=-m_P^2$, with $m_P>0$. Then, assuming that $m_P <
M_f + M_{f'}$, the magnetic contribution can be written as (see
App.~B)
\begin{align}
c^{mag}_{ff'} (0,q_\parallel^2=-m_P^2) = 2 N_c \left\{ \dfrac{
I^{mag}_{1f} + I^{mag}_{1f'}}{2} - \left[ m_P^2 - (M_f - M_{f'})^2
\right] I^{mag}_{2ff'}(-m_P^2)  \right\}\, . \label{cmagff'}
\end{align}
The function $I^{mag}_{1f}$ has already been expressed in Eq.~\eqref{i1} while
\begin{align}
I_{2ff'}^{mag}(-m_P^2) = \dfrac{1}{8\pi^2}
\,\lim_{\epsilon\rightarrow0}\, \int_{0}^{1} dy \left[ \psi(\bar
x_{ff'}-i\epsilon) - \ln(\bar x_{ff'}-i\epsilon) + \dfrac{1}{2
(\bar x_{ff'}-i\epsilon)} \right]\, , \label{I2ff'mag1}
\end{align}
where $\psi(x)$ is the digamma function and we have defined
\begin{align}
\bar x_{ff'} = \dfrac{yM_f^2 +(1-y)M_{f'}^2 -y(1-y)m_P^2}{2B_f} \, . %\label{xff'}
\end{align}

For $m_P < M_f + M_{f'}$ we have that $\bar x_{ff'} >0$ for all
values of $y$ within the integration range of the integral of
Eq.~\eqref{I2ff'mag1}. Thus, the limit $\epsilon\rightarrow 0$ can
be directly taken.

On the other hand, for $\eta'$ we expect that $m_P > M_f +
M_{f'}$. In this case  one has to have special care since $\bar
x_{ff'}$ can be negative within the interval $0 < y < 1$. We
proceed by taking the analytic continuation of both the digamma
and logarithm functions. This implies that the inverse propagators
become complex functions. Thus, we assume that $q_\parallel$
develops an imaginary part
\begin{equation}
q_\parallel^2 = -\left( m_P - \dfrac{i}{2} \Gamma_P \right)^2 \, ,
\end{equation}
where $\Gamma_P$ is associated with the decay width of the meson.
Following the customary method introduced in Ref.~\cite{Rehberg:1995kh}, we assume that the width is not too large and neglect its contribution
inside $I^{mag}_{2ff'}$ function (this also applies to the equivalent vacuum contribution)
\begin{equation}
c^{mag}_{ff'} (m_P,\Gamma_P) \simeq 2 N_c \left\{ \dfrac{
I^{mag}_{1f} + I^{mag}_{1f'}}{2} - \left[ \left(m_P - \dfrac{i}{2}
\Gamma_P \right)^2 - (M_f - M_{f'})^2 \right]
I^{mag}_{2ff'}(-m_P^2)  \right\} \, .
\end{equation}
Note that in Eq.~\eqref{I2ff'mag1} one might hit some poles of the
digamma function if the limit $\epsilon\rightarrow 0$ is naively
taken. As detailed in App.~B, through a careful treatment of these
poles one can explicitly calculate the $ I^{mag}_{2ff'}$ function.
The general result for $f \neq f'$ is given in Eq.~\eqref{I2BBfin}
of App.~B. We remark here that, as a consistency check, we have
repeated the calculation using the Landau level representation of
the quark propagator, well-defined for all
$m_P$, obtaining the same result. For the determination of the
$\eta'$ mass we only need the $f=f'$ version of the general
expression, given by
\begin{align}
I_{2ff}^{mag}(-m_P^2) =& -\dfrac{1}{8\pi^2} \Bigg\{
\ln\left(\dfrac{M_f^2}{2 B_f}\right)  + 2\beta_0 \ln \left[\dfrac{
m_P(1+\beta_0)}{ 2M_f } \right] - 2
+  \dfrac{2B_f }{m_P^2}\sum_{n=0}^N \dfrac{ g_n }{\beta_n} \ln \left(\dfrac{1-\beta_n}{1+\beta_n}\right) \Bigg\} \nonumber\\[2mm]
& + \dfrac{1}{8\pi^2} \int_{0}^{1} dy  \ \psi(\bar x_{ff} + N + 1)
+ \dfrac{i}{8\pi} \left[ \beta_0 -\dfrac{2B_f}{m_P^2} \sum_{n=0}^N
\dfrac{g_n}{\beta_n} \right]\, ,
\end{align}
where $g_n= 2 - \delta_{n0}$ and $N= \mathrm{Floor} \left[ m_P^2\beta_0^2/8 B_f \right]$. Moreover,
\begin{equation}
\beta_n = \sqrt{ 1 - \dfrac{4M_f^2}{m_P^2} - \dfrac{8nB_f}{m_P^2}
}\, .
\end{equation}

For the neutral kaons, we expect $m_{K^0}=m_{\bar K^0}<M_d+M_s$. In this case the polarization function is real and $I^{mag}_{2ff'}$ is well defined in the
$\epsilon\rightarrow 0$ limit of Eq.~\eqref{I2ff'mag1}.
Therefore, the pole-mass will be given by the solution of
\begin{equation}
\mathcal{G}_{K^0}(q_\perp^2=0,q_\parallel^2=-m_{K^0}^2) = 0 \, .
\end{equation}

In the $P, P'=\pi_3,\eta_0,\eta_8$ subspace,
the corresponding quadratic action can be expressed in matrix
notation through the following inverse matrix propagator
\begin{align}
\mathcal{M} = \begin{pmatrix}
\mathcal{G}_{\pi_3\pi_3} & \mathcal{G}_{\pi_3\eta_0} & \mathcal{G}_{\pi_3\eta_8} \\
\mathcal{G}_{\eta_0\pi_3} & \mathcal{G}_{\eta_0\eta_0} & \mathcal{G}_{\eta_0\eta_8} \\
\mathcal{G}_{\eta_8\pi_3} & \mathcal{G}_{\eta_0\eta_8} & \mathcal{G}_{\eta_8\eta_8}
\end{pmatrix} \, ,
\label{Mprop}
\end{align}
which is actually symmetric. The physical meson pole-masses and widths will be given by the roots of
\begin{equation}
\det[\mathcal{M}(m_P,\Gamma_P)] = 0 \, , \label{eqnneutral}
\end{equation}
where the three pair of roots are to be associated with the $\pi^0,\eta,\eta'$. Of course, one expects to
get $\Gamma_{\pi^0}=\Gamma_{\eta}=0$ while $\Gamma_{\eta'}$ is expected to be non-vanishing.
Note that when $B=0$, $\pi_3$ (and therefore $\pi^0$) decouples from the $\eta_0,\eta_8$ states due to isospin symmetry. However, in the presence of an external magnetic field this symmetry breaks down due to different quark electric charges. In this case, the $\pi^0,\eta,\eta'$ neutral mesons consist of a mix of $\pi_3,\eta_0,\eta_8$ states, reflected by the fact that non-diagonal terms are present in the inverse propagator of Eq.~\eqref{Mprop}.

\vspace{8mm}
\subsubsection{Charged mesons}

In this case the contributions of Schwinger phases associated with the quark
propagators do not cancel out, leading to a breakdown of translational invariance.
In order to diagonalize the charged meson fields, we employ the Ritus-like formalism.
We find it convenient to introduce the following notation convention
\begin{equation}
\delta P(x) \, = \, \sumint_{\bar{q}}\mathbb{F}^P_{\bar{q}}(x) \
\delta P(\bar{q}) \: ; \qquad \sumint_{\bar{q}} \, \equiv \,
\dfrac{1}{(2\pi)^4} \sum_{k=0}^\infty \int dq_2 \;dq_3\; dq_4 \, ,
\end{equation}
with $\bar{q}=(k,q_2,q_3,q_4)$ where $k$ labels the charged meson Landau level. The Ritus-like eigenfunctions are
\begin{align}
\mathbb{F}^P_{\bar{q}}(x) = N_k e^{i(q_2x_2+q_3x_3+q_4x_4)}
D_k(\rho_P)\, .
\end{align}
Here $D_k(x)$ are the cylindrical parabolic functions. We have
also defined $N_k=(4\pi B_P)^{1/4}/\sqrt{k!}$ and
$\rho_P=\sqrt{2B_P}\ x_1-s_P\sqrt{2/B_P}\ q_2$, where  $B_P=|Q_P
B|$ and $s_P=\mathrm{sign}(Q_P B)$. Note that in our case, for the
$\pi^\pm$ and $K^\pm$ mesons  these definitions reduce to
$B_P=|eB|$ and $s_{\pi^\pm}=s_{K^\pm}=\pm 1$.

The corresponding transformed polarization functions will be diagonal in $\bar{q}, \bar{q}'$ space.
Thus, the charged meson contribution to the quadratic action in the Ritus basis can be written as
\begin{align}
S^{\,\mbox{\tiny quad}}_{char.mes} \, = \, \dfrac{1}{2}
\sumint_{\bar{q}} \, \sum_{P=\pi^\pm, K^\pm} \, \delta
P^*(\bar{q})\ {\cal G}_{P}(k,\Pi^2)\ \delta P(\bar{q})\, ,
\end{align}
where $\Pi^2 = q_\parallel^2 + (2k + 1)B_P$. The inverse
propagators read
\begin{align}
{\cal G}_{\pi^\pm}(k,\Pi^2) &= \left[ 2 G - K \phi_s \right]^{-1} - c_{ud}(k,\Pi^2)\, , \nonumber\\
{\cal G}_{K^\pm}(k,\Pi^2) &= \left[ 2 G - K \phi_d \right]^{-1} - c_{us}(k,\Pi^2)\, ,
\end{align}
where we have used
\begin{align}
\int d^4x' \, d^4x \:
\left[ \mathbb{F}^{\pi^+}_{\bar{q}}(x) \right]^\ast c_{ud}(x,x') \: \mathbb{F}^{\pi^+}_{\bar{q}'}(x')
&= c_{ud}(k,\Pi^2) \, (2\pi)^4 \delta_{\bar{q},\bar{q}'} \,   , \nonumber\\[2mm]
\int d^4x' \, d^4x \:
\left[ \mathbb{F}^{K^+}_{\bar{q}}(x) \right]^\ast c_{us}(x,x') \: \mathbb{F}^{K^+}_{\bar{q}'}(x') &
= c_{us}(k,\Pi^2) \, (2\pi)^4 \delta_{\bar{q},\bar{q}'} \,.
\end{align}
These functions are divergent and need to be regularized. Within the MFIR scheme using a 3D cutoff they can be expressed as
\begin{align}
c_{ff'}^{reg}(k,\Pi^2) \, = \, c_{ff'}^{vac\vphantom{g}}(\Pi^2) + c_{ff'}^{mag}(k,\Pi^2)\, . \label{cff'carg}
\end{align}
The regularized vacuum contribution is given in App.~A. After a long but straightforward calculation (see~\cite{Coppola:2019uyr} for details), we obtain the following expression for the magnetic contribution
\begin{align}
c_{ff'}^{mag}(k,\Pi^2)  \, =& \,  \dfrac{N_c}{2\pi^2} \int_0^\infty dz \int_0^1 dy \,
e^{-z\left[ yM_f^2 + (1-y)M_{f'}^2 + y(1-y)\Pi^2 \right]} \nonumber\\[2mm]
& \times \bigg\{ \left[ M_f M_{f'} + \dfrac{1}{z} - y(1-y)\left(\Pi^2 - (2k+1)B_P \right) \right] \nonumber\\[2mm]
& \hphantom{\times} \times \left[ \dfrac{(1 + s_f s_{f'} t_f t_{f'})}{\alpha_+} \left( \dfrac{\alpha_-}{\alpha_+} \right)^k
\, e^{zy(1-y)(2k+1)B_P} -\dfrac{1}{z} \right] \nonumber\\[2mm]
& \hphantom{\times} + \dfrac{(1-t_f^2)(1-t_{f'}^2)}{\alpha_+^2 \alpha_-} \left( \dfrac{\alpha_-}{\alpha_+} \right)^k
\left[ \alpha_- + k(\alpha_- - \alpha_+) \right]  \, e^{zy(1-y)(2k+1)B_P}  \nonumber\\[2mm]
& \hphantom{\times} - \dfrac{1}{z}\left[\dfrac{1}{z}
-y(1-y)(2k+1)B_P \right] \bigg\} \, , \label{cff'magcar}
\end{align}
where we have introduced the definitions $t_f=\tanh(B_fzy)$,
$t_{f'}=\tanh[B_{f'}z(1-y)]$ and $\alpha_\pm=(B_{f'}t_f + B_f
t_{f'} \pm B_P t_f t_{f'})/(B_f B_{f'})$. Note also that
$Q_P=Q_f-Q_{f'}$.

For pions and kaons we expect them to develop only a real
pole-mass, i.e. $\Pi^2=-m_P^2$ with $m_P<M_f+M_{f'}$. In that case
the integrals in Eq.~\eqref{cff'magcar} are convergent and
well-defined. Therefore, for each Landau level the charged mesons
pole-masses will be given by the solutions of
\begin{align}
\mathcal{G}_{\pi^\pm}(k,\Pi^2=-m_{\pi^\pm}^2) &= 0 \, , \nonumber\\[2mm]
\mathcal{G}_{K^\pm}(k,\Pi^2=-m_{K^\pm}^2) &= 0 \, .
\end{align}

\section{Numerical results}

To obtain numerical results for the magnetic field dependence of the meson masses one has to fix the model parametrization. Here,
following
Ref.~\cite{Rehberg:1995kh}, we take the parameter set $m_u = m_d = 5.5\ \mbox{MeV}$, $m_s = 140.7\ \mbox{MeV}$,
$\Lambda = 602.3 \ \mbox{MeV}$, $G\Lambda^2 = 1.835$ and $K\Lambda^5= 12.36$, which
has been determined on fixing that for vanishing external field one gets
$m_\pi = 135 \ \mbox{MeV}$, $m_K = 497.7 \ \mbox{MeV}$,
$m_{\eta'} = 957.8 \ \mbox{MeV}$ and $f_\pi = 92.4 \ \mbox{MeV}$.
This parameter set gives an $\eta$ mass of $m_\eta = 514.8\ \mbox{MeV}$,
which compares reasonably well with the physical value $m^{phs}_\eta = 548.8 \ \mbox{MeV}$, together with an appropriate value for the chiral condensate of $\langle \bar\psi_f \psi_f \rangle^{1/3}=242$ MeV for $f=u,d$.
As mentioned in the Introduction, while local NJL-like models are able to reproduce the MC effect
at vanishing temperature, they fail to lead to the IMC effect.
Among the possible ways to deal with this problem, one of the simplest consists of allowing the model parameters to depend on the magnetic field.
Motivated by this we also explore the possibility of considering a magnetic field dependent coupling $G(B)$. We adopt the one proposed in
Ref.~\cite{Ferreira:2014kpa} in the context of an SU(3) NJL model with the same parameters that we use.
In that work the current quark masses, $\Lambda$ and $K$ were kept constant while for $G(B)$ the form
\begin{equation}
G(B) = G \, \left[ \dfrac{ 1 + a (eB/\Lambda^2_{QCD})^2 + b (eB/\Lambda^2_{QCD})^3 }{ 1 + c (eB/\Lambda^2_{QCD})^2 + d (eB/\Lambda^2_{QCD})^4 } \right] \, , \label{fFCLFP}
\end{equation}
was introduced. Here, $a=0.0108805$, $b=-1.0133 \ 10^{-4}$, $c=0.02228$, $d=1.84558 \ 10^{-4}$ and $\Lambda_{QCD} = 300 \ \mbox{MeV}$.
As stated in Ref.~\cite{Ferreira:2014kpa}, this form of the scalar coupling has been fitted so that the lattice QCD pseudocritical chiral
transition temperatures are reproduced.

\begin{figure}[tb]
    \centering{}\includegraphics[width=0.6\textwidth]{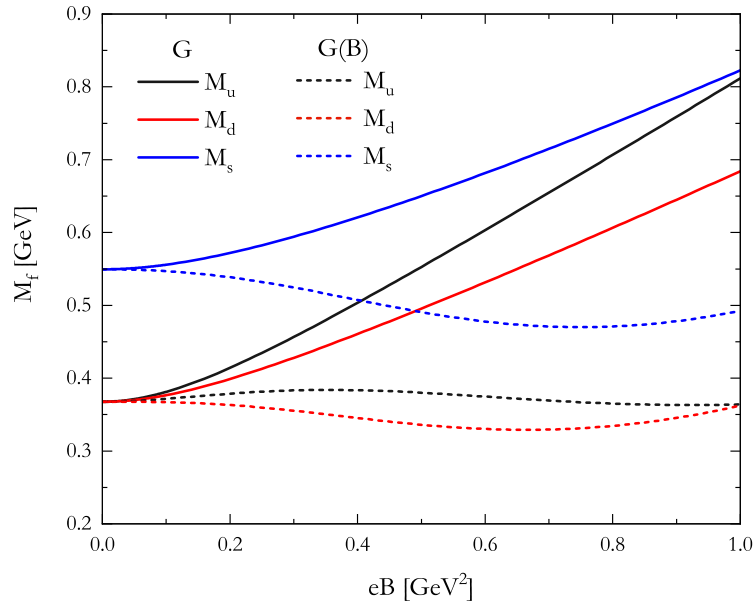}
\caption{(Color online) Effective quark masses $M_u$ (black), $M_d$ (red) and $M_s$ (blue) as functions of $eB$ for fixed (solid lines) and $B$-dependent (dashed lines) coupling $G$.}
\label{Fig1}
\end{figure}

Results for the magnetic field dependence of the dynamical quark masses are shown in Fig.~\ref{Fig1},  for both constant and $B$-dependent coupling $G$.
As we see, for constant $G$ all quark masses increase with $B$.
 In contrast, for $G(B)$ they display a non-monotonous behavior, less affected by the magnetic field. In this case, $M_d$ and $M_s$ initially decrease with $B$, while about $eB \sim 0.6-0.7$ GeV$^2$
this tendency reverses. On the other hand, $M_u$ has just the opposite behavior.  In fact, these dependencies of the dynamical
quark masses on the magnetic field are roughly consistent with
the results obtained in Ref.~\cite{Endrodi:2019whh}. In that work these quantities
have been extracted from a LQCD calculation of the baryon masses using a simple minded approximation based on the
constituent quark model.

\begin{figure}[tb]
    \centering{}\includegraphics[width=0.75\textwidth]{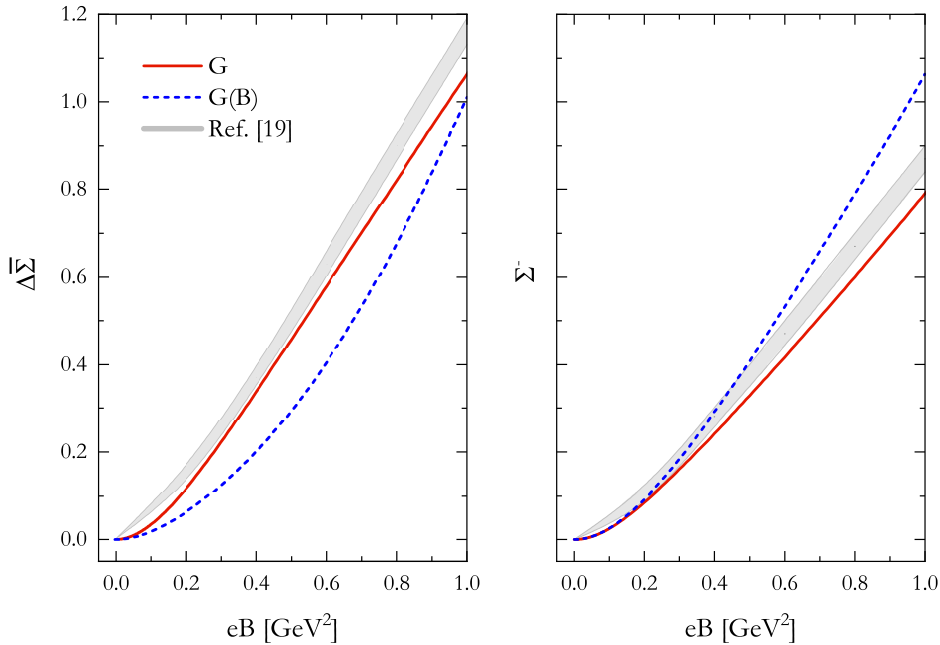}
\caption{(Color online) Left: average condensate as a function of
$eB$. Right: condensate difference as a function of $eB$. Results for constant (red solid lines) and  $B$-dependent (blue dashed lines) coupling $G$ are shown.  LQCD results from Ref.~\cite{Bali:2012zg} (gray bands) are added for comparison.}
\label{Fig2}
\end{figure}
It should be stressed that in spite of the rather different behavior between the dynamical quark masses, a magnetic catalysis effect at zero temperature
is obtained independently on whether $G$ depends on $B$ or not. This is shown in Fig.~\ref{Fig2}, where we displayed the conveniently normalized light quark
condensates. These quantities are defined as it follows. As in Ref.~\cite{Bali:2012zg}, for the case of vanishing temperature
we are interested in, we first introduce
\begin{equation}
\Sigma_f = \dfrac{2 m_f}{D^4} \left[ \phi_f^{reg}(B) - \phi_f^{reg}(0) \right] + 1 \, ,
\label{Sig}
\end{equation}
where we have explicitly stated the magnetic field dependence of the quark condensate, defined in Eq.~\eqref{phif}.
Moreover, $D = (86 \ \mbox{MeV} \times 135 \ \mbox{MeV})^{1/2}$ was introduced in Ref.~\cite{Bali:2012zg} as a kind of normalization constant and
$m_f$ is the current quark mass of each light flavor.
%We have added an upper index to distinguish the $B=0$ case from the one corresponding to finite B.
Then, in the left panel of Fig.~\ref{Fig2}
we plot $\Delta \bar \Sigma = (\Sigma_u + \Sigma_d)/2 -1$ while in the right panel the difference $\Sigma^- = \Sigma_u-\Sigma_d$ is shown.
The gray bands in Fig.~\ref{Fig2} correspond to LQCD results taken from Ref.~\cite{Bali:2012zg}, whereas full red (dashed blue) lines
represent our results for constant $G$ ($B$-dependent $G$). We observe that although the predictions for constant $G$ are somewhat closer to the
LQCD results, those corresponding to $G(B)$ can certainly be considered as acceptable. It is interesting to remark here that other form functions of $G(B)$, such as the ones proposed in Refs.~\cite{Endrodi:2019whh,Avancini:2016fgq}, reproduce similar trends for these quantities.

\begin{figure}[tb]
    \centering{}\includegraphics[width=0.75\textwidth]{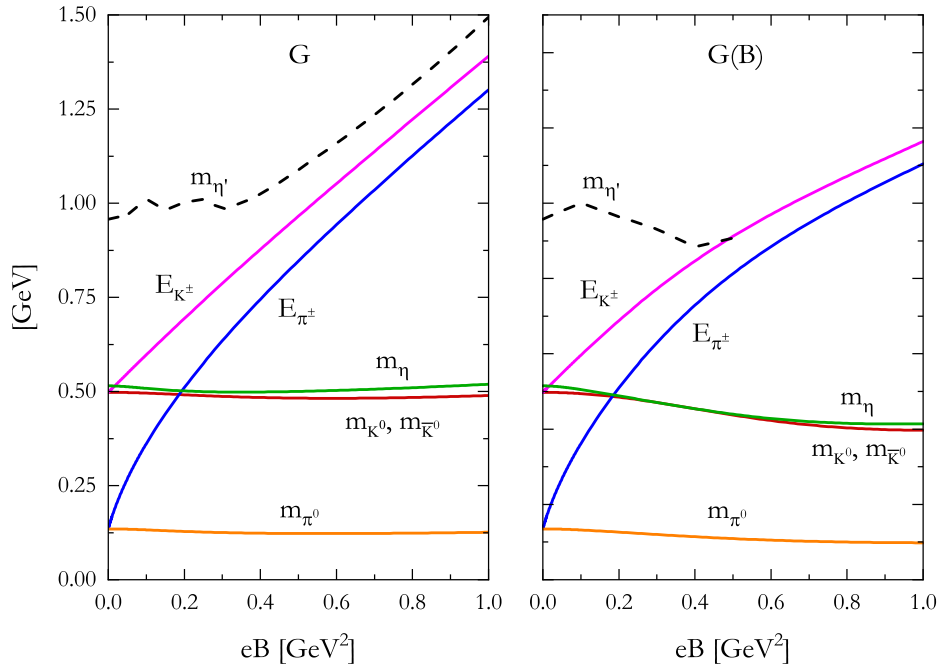}
\caption{(Color online) Pseudoscalar neutral meson pole-masses and charged mesons lowest energies as functions of $eB$ for constant (left) and $B$-dependent (right) coupling $G$.}
\label{Fig3}
\end{figure}

We turn now to our results for the magnetic field dependence of the masses of the nonet of pseudoscalar mesons. They are shown in Fig.~\ref{Fig3}, where for charged mesons we instead display their lowest energy states, given by
\begin{equation}
E_{P^\pm} = \sqrt{m_{P^\pm}^2 + (2k+1)eB + q_3^2} \, \bigg\rvert_{\substack{q_3=0 \\ k=0}} = \sqrt{m_{P^\pm}^2 + eB} \, ,
\end{equation}
(note that both $E_P$ and $m_P$ depend on $B$ although not
explicitly stated). The left (right) panel corresponds to the case of
constant coupling $G$ ($B$-dependent $G$). We observe that, except
for the $\eta'$-mass, the $B$-dependence is rather mild in the
case of the neutral mesons. On the other hand a rather strong
increase with growing $B$ is found for charged meson masses. These
results are analyzed in further detail in what follows.

The case of $\eta'$ is somewhat special and, therefore, indicated in dashed lines in Fig.~\ref{Fig3}. In fact, already at $B=0$ its mass is above the threshold for
$q \bar q$-decay and, thus, the associated $q \bar q$ polarization diagram receives an unphysical imaginary part. Following Ref.~\cite{Rehberg:1995kh}
we accept this as an unavoidable feature of the NJL model and define the $\eta'$-mass as the real part of the corresponding pole in the complex plane.
We should keep in mind, however, that this fact makes the predictions for the $\eta'$-mass less reliable as compared to those of the
other mesons. The situation worsens for finite magnetic field. First, new divergencies appear at low magnetic fields due the
existence of thresholds associated with the Landau levels of the intermediate quark states. Although these divergencies are along the real axis,
they originate the kind of oscillatory behavior found for $eB \lesssim 0.2 \ \mbox{GeV}^2$. In passing, we note that including in the calculation the imaginary part of the polarization function makes these divergencies less harmful. If one neglects that
contribution, as done in Ref.~\cite{Hatsuda1994}, the determination of $m_{\eta'}$ becomes full of ambiguities making its determination
even more troublesome. The other point has to do with the fact that at finite magnetic field the width is in general larger than the
already non-negligible value at $B=0$, $\Gamma^{B=0}_{\eta'}= 269 \ \mbox{MeV}$. 
For constant $G$, we encounter a nonmonotonic behavior of the width, which shows a close-to-vacuum mean value of $\Gamma^{B,mean}_{\eta'}= 332 \ \mbox{MeV}$ but can reach values of $\Gamma^{B}_{\eta'} \sim 590 \ \mbox{MeV}$ at intermediate fields.
On the other hand, for $B$-dependent $G$ the pace of growth of the width increases. At fields strengths around $eB\sim 0.5 \ \mbox{GeV}^2$ the width exceeds the mass, with a value of $\Gamma^{B}_{\eta'} \sim 1.46 \ \mbox{GeV}$. 
This enhancement of the width, together with the decrease of $G(B)$ as $B$ increases, results in the fact that for $eB \gtrsim 0.5 \ \mbox{GeV}^2$
no solution of Eq.~\eqref{eqnneutral} can be found apart from the ones associated with $\pi^0$ and $\eta$. Namely, above such a value of the magnetic field the coupling strength is not enough to form an $\eta'$-resonance
in the $q \bar q$-continuum.

\begin{figure}[tb]
    \centering{}\includegraphics[width=.8\textwidth]{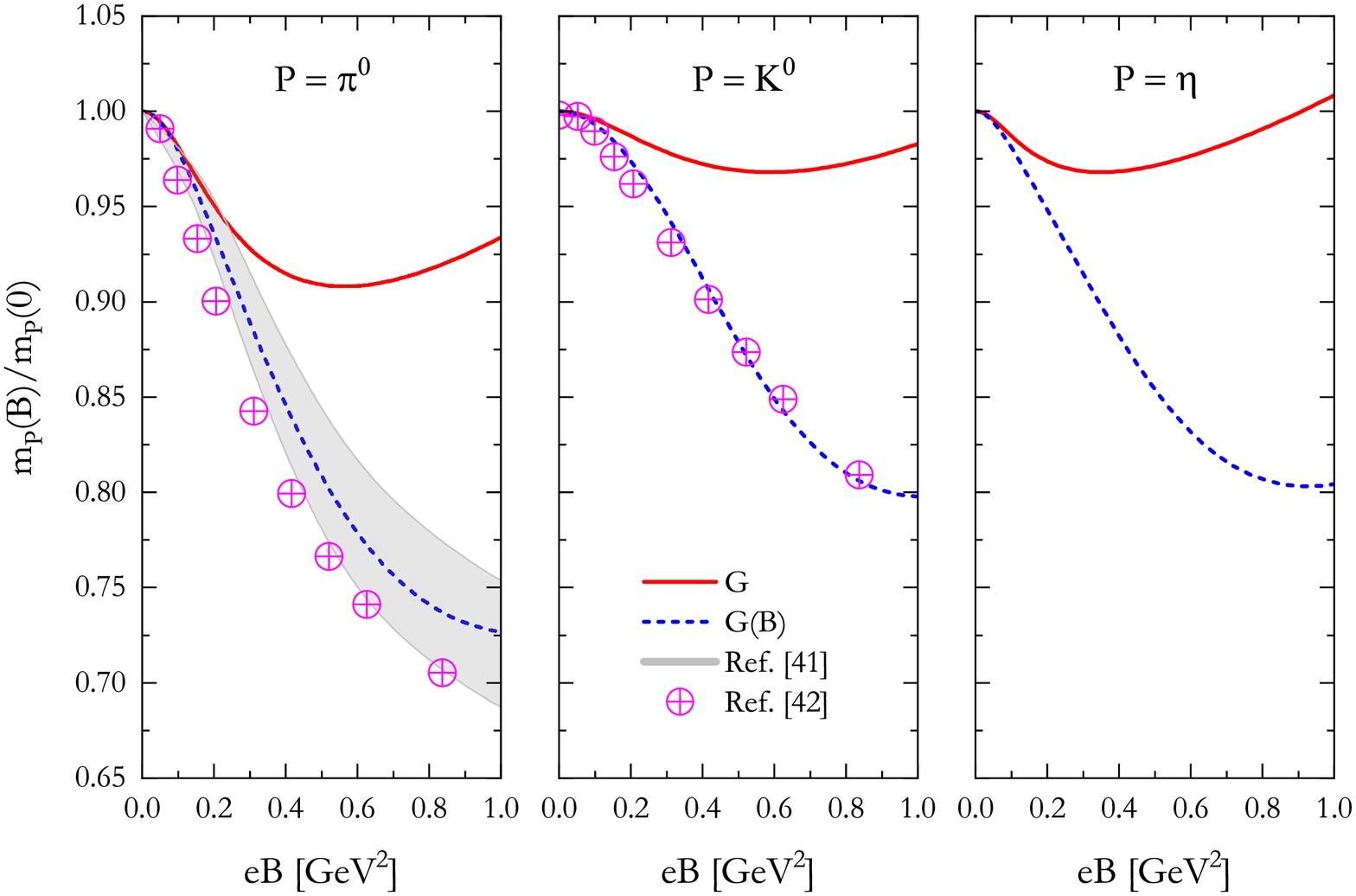}
\caption{(Color online) Normalized neutral meson masses as functions of $eB$ for constant (red solid lines) and $B$-dependent (blue dashed lines) coupling $G$. LQCD results from Ref.~\cite{Bali:2017ian}  (grey band) and Ref.~\cite{Ding2021}  (magenta circles) are added for comparison.}
\label{Fig4}
\end{figure}

To discuss our results for the other neutral mesons ($\pi^0$, $K^0,\bar K^0$ and $\eta$) in more detail we display in Fig.~\ref{Fig4}
the corresponding  masses taken with respect to their values at $B=0$. We show results using a constant and a $B$-dependent coupling $G$ together with LCQD simulations from Refs.~\cite{Bali:2017ian,Ding2021} for comparison.
It should be noticed that
these LQCD calculations correspond to non-physical pion masses i.e. $415$ and $220 \ \mbox{MeV}$, respectively, for vanishing magnetic field.
In both cases they point to a stronger decrease of the $\pi^0$ mass with increasing $B$ than the one found in our calculation with constant $G$.
On the other hand, the results obtained using a $B$-dependent $G$ are in reasonable good agreement with LQCD ones. A similar
observation have been made in Ref.~\cite{Avancini:2016fgq} in the context of a two-flavor NJL model. This seems to also provide further support
to the relation between the IMC effect and the reduction of the neutral pion mass at finite $B$ mentioned in Ref.~\cite{Ding2021}. In the
case of $K^0$ and $\bar K^0$ masses (central panel), the only LQCD result that has been reported is that of Ref.~\cite{Ding2021}.
We observe that, once again, a much better agreement with these results are obtained when a $B$-dependent coupling $G$ is used in the NJL model.
Finally, in the right panel we show our predictions for the behavior of the normalized $\eta$-meson mass. They turn out to be quite similar
to the ones obtained for the $K^0$ and $\bar K^0$ relative masses.

\begin{figure}[tb]
    \centering{}\includegraphics[width=0.92\textwidth]{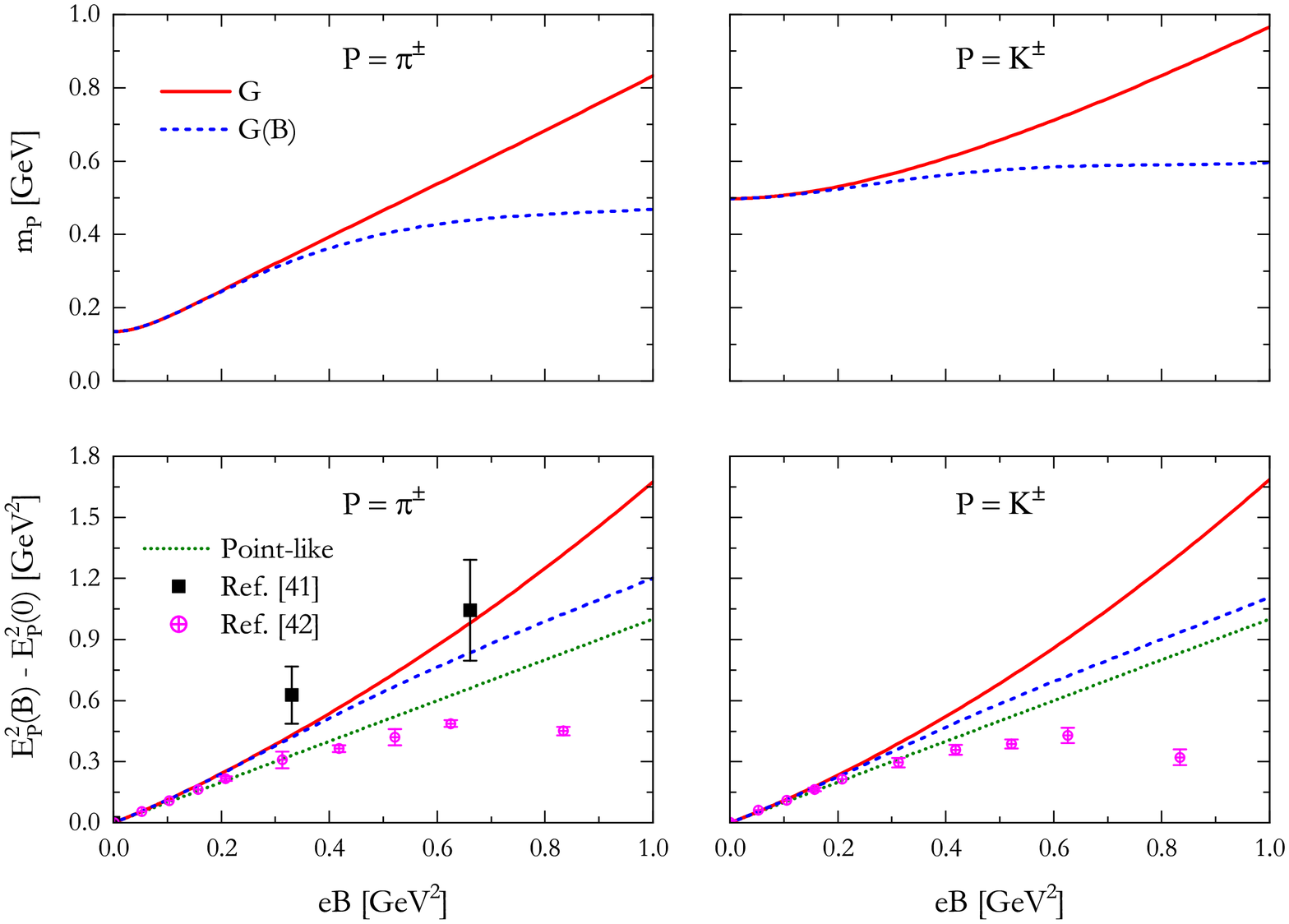}
\caption{(Color online) Charged meson masses (top) and differences of squared lowest energies between the case at $B\neq 0$ and $B=0$ (bottom) for charged pions (left) and kaons (right) as a function of $eB$. Results for constant and $B$-dependent coupling $G$ are shown in red solid and blue dashed lines, respectively. Green dotted lines correspond to energies associated with point-like charged mesons. LQCD results from 
Ref.~\cite{Bali:2017ian} (black squares) and Ref.~\cite{Ding2021}  (magenta circles) are added for comparison.}
\label{Fig5}
\end{figure}

Finally, we consider the masses of charged pseudoscalar mesons $\pi^\pm$ and $K^\pm$. In Fig.~\ref{Fig5} we display the differences in their squared lowest energies from the case of a zero magnetic field, i.e. $E^2(B) - E^2(B=0)$. We also include their masses in the top graphs for completeness. We show results for $G$ and $G(B)$ as compared to a point-like charged meson and LQCD simulations from Refs.~\cite{Bali:2017ian,Ding2021}.
We observe
that for both charged pion and kaons our results show a stronger increase with growing $B$ as compared with the ones associated with
point-like mesons. Those obtained using a $B$-dependent $G$ are, however, somewhat closer to them. As for the comparison with LQCD results
we note that in the case of charged pions there are significant differences between the results reported by the two different LQCD groups, specially at large magnetic fields. Although our results seem to be more consistent with those of Ref.~\cite{Bali:2017ian} it should
be recalled that they correspond to a larger (unphysical) value of the $B=0$ pion mass and have larger error bars. In any case, we see
that, for both charged pions and kaons, our NJL results show no sign of the strong non-monotonous behavior found in the LQCD calculation of
Ref.~\cite{Ding2021}.
Results obtained within the SU(2)
version of the model~\cite{Xu:2020yag}, seem to indicate that the
inclusion of quarks anomalous magnetic moments does not modify the
trend of the charged pion mass obtained in the present work.

\section{Conclusions}

In this work we have considered the masses of the light
pseudoscalar masses under the influence of strong magnetic fields
in the framework of the SU(3) Nambu--Jona-Lasinio model that
includes the 't Hooft-Maekawa flavor mixing interaction. The model
parameters have been determined on fixing that for vanishing
external field one reproduces the physical values of the $\pi$,
$K$ and $\eta'$ meson masses together with the
pion weak  decay constant.
The possibility of using
a magnetic field dependent four-fermion coupling constant in order
to reproduce the inverse magnetic catalysis at finite temperature
has also been considered. Since the NJL model is not
renormalizable, the calculation of observables requires an
appropriate regularization scheme in order to deal with
ultraviolet divergences. Here we have used the magnetic field
independent regularization procedure, in which only divergent
vacuum contributions to quantities at zero external magnetic field
are regularized. This scheme has been shown to provide more
reliable predictions in comparison with other regularization
methods often used in the literature~\cite{Avancini:2019wed}.

At the mean field level, effective quark masses, shown in Fig.~\ref{Fig1}, steadily increase with $B$ for constant $G$ but display a non-monotonous behavior for $G(B)$ which resembles the one found in Ref.~\cite{Endrodi:2019zrl}. Moreover, our results for the difference and average of the condensates calculated for both constant $G$ and $G(B)$ and their comparisons with the available LQCD results, as shown in Fig.~\ref{Fig2}, support the fact that the parametrizations used in this paper are in a very reasonable agreement with LQCD results.

In order to study meson masses we go beyond the mean-field approximation, considering second order corrections to the bosonized Euclidean action of the SU(3) NJL model. Mesons are treated as quantum fluctuations in the random
phase approximation. While for neutral mesons one can take the usual momentum basis to diagonalize the corresponding polarization functions, this is not possible for charged mesons since Schwinger phases do not cancel out. In that case, we have employed a method based on the Ritus eigenfunction approach to magnetized relativistic systems.
As discussed in Sec.~II,  at the quadratic level the inverse
propagators corresponding to the neutral $\pi_3$, $\eta_0$ and
$\eta_8$ fields  are arranged in terms of a symmetric 3x3 matrix;
the pole-masses and widths of the
physical mesons ($\pi^0$, $\eta$ and $\eta^\prime$) are obtained
as the roots of this inverse propagator matrix determinant. Note
that in the $B \ne 0$ case, besides the 't Hooft-Maekawa
interaction which breaks the $U_A(1)$ symmetry and is responsible
for the coupling between the $\eta_0$ and $\eta_8$, the magnetic
field also breaks the isospin symmetry, leading to
a mixing between all three states~\cite{Cao2021}. This is in contrast to the
$B=0$ case where due to the isospin symmetry ($M_u=M_d$)
the $\pi_3$ field is decoupled leaving only
$\eta_0$ and $\eta_8$ mixed in a symmetric 2x2 matrix.

As already known from the usual SU(3) NJL model at $B=0$, the
$\eta^\prime$ meson comes out in the model as a resonance or
unstable particle. In this case, the propagator becomes a complex
number and from the analysis of the complex pole, the mass
of the resonance is
obtained. In the presence of a finite magnetic field, the
situation is more dramatic since the propagators may develop
several poles depending on $B$, which have to be properly treated.
We have developed in this paper a new formalism to deal with this
situation. Of course, the results for the $\eta^\prime$ meson are
less reliable and its calculation certainly approaches the limit
of applicability of the NJL model, since this model does not
include confinement. In fact, we find that
using $G(B)$ the coupling strength is not enough to form an
$\eta^\prime$-resonance for  $eB \gtrsim 0.5 \ \mbox{GeV}^2$.

Our results for the normalized $\pi^0$ mass in Fig.~\ref{Fig4} show that,
for constant $G$, the mass displays a
non-monotonous behavior with $B$, which initially decreases but is
afterwards enhanced for $eB \gtrsim 0.5 \ \mbox{GeV}^2$. On the
other hand, using a $B$-dependent coupling $G(B)$ we recover the
monotonous decreasing behavior found in LQCD results. Something
similar happens with $K^0$ and $\bar K^0$ masses. For $\eta$, our
prediction is similar to that of $K^0$ and $\bar K^0$. We thus
conclude that incorporating the inverse magnetic catalysis in the
NJL model, here through the $G(B)$ coupling,  is fundamental for
qualitatively reproduce the available LQCD results.
Concerning charged mesons, our results for
the differences in their squared lowest energies from the $B=0$
case are shown in Fig.~\ref{Fig5}, where a strong enhancement
with $B$ is seen. This increase even surpasses the one associated
with a point-like charged meson. Our NJL results are in reasonable
agreement with LQCD results of Ref.~\cite{Bali:2017ian} within
error bars  On the other hand, no sign of
the non-monotonous behavior found in the LQCD calculation of
Ref.~\cite{Ding2021} is observed.

\acknowledgments

This work has been supported in part by Consejo Nacional de Investigaciones
Cient\'ificas y T\'ecnicas and Agencia Nacional de Promoci\'on Cient\'ifica
y Tecnol\'ogica (Argentina), under Grants No.~PIP17-700 and No.~PICT17-03-0571 respectively;
by Conselho Nacional de Desenvolvimento Cient\'ifico
e Tecnol\'ogico  (CNPq), Grant No. 304518/2019-0 (S.S.A.); 
by Coordena\c{c}\~ao de Aperfei\c{c}oamento de Pessoal
de N\'ivel Superior (CAPES-Brazil) - Finance Code 001 (J.C.S.);
and also is part of the project Instituto Nacional de Ci\^encia
e Tecnologia - F\'isica Nuclear e Aplica\c{c}\~oes (INCT - FNA), Grant No.~464898/2014-5 (S.S.A. and J.C.S.).
\vfill
\pagebreak

%%%%%%%%%%%%%%%%%%%%%%%%%%%%%%%%%%%%%%%%%%%%%%%

\appendix
\section{Explicit expression of the vacuum function $\mathbf{c^{vac}_{ff'}(q^2)}$}

The functions $c^{vac}_{ff'}$ appear in
Eqs.~\eqref{cff'neutro} and~\eqref{cff'carg}. In their unregularized form they are defined by
\begin{align}
c^0_{ff'} (q^2) = 2 N_c \int_p tr_D \left[ \tilde{S}^{f,0}_{p_-} \ \gamma_5 \  \tilde{S}^{f',0}_{p_+}\ \gamma_5 \right] \, ,
\end{align}
where $\tilde{S}^{f,0}_{p}=1/(\, \rlap/p+M_f)$ is the usual vacuum propagator for a quark of mass $M_f$.
Here, $p_\pm = p \pm q/2$. We recall that in this work all four-momenta are defined in Euclidean
space. By taking the trace and integrating over $p$ one obtains
\begin{align}
c^0_{ff'} (q^2) \: = \ & \dfrac{N_c}{2\pi^2} \int_0^\infty dz \int_0^1 \: dy
\exp\left\{-z\left[ y M_f^2 + (1-y) M_{f'}^2 + y(1-y) q^2 -i\epsilon \right] \right\} \times \nonumber\\[2mm]
& \qquad \qquad \qquad \qquad \dfrac{1}{z} \left[ M_f M_{f'} + \dfrac{2}{z} - y(1-y)q^2 \right] \, . \label{c0ff'PT}
\end{align}

We have expressed this function in the proper time formalism. Through some algebraic manipulation, it can also be written in the following standard form
\begin{align}
c^0_{ff'} (q^2) = 2 N_c \left\{ \dfrac{ I^0_{1f} + I^0_{1f'}}{2} + \left[ q^2 + (M_f - M_{f'})^2 \right] I^0_{2ff'}(q^2) \right\}\, ,  \label{c0ff'}
\end{align}
where the integrals $I^0_{1f}$ and $I^0_{2ff'}$ are defined by
\begin{align}
I^0_{1f}  &= 4 \int_p \: \dfrac{1}{p^2 + M_f^2} \, ,\nonumber\\[2mm]
I^0_{2ff'}(q^2) &= -2 \int_p \: \dfrac{1}{(p_-^2 + M_f^2 - i\epsilon)(p_+^2 + M_{f'}^2 - i\epsilon)} \, .
\end{align}

In order to  regularize the vacuum loop integrals we introduce a 3D cutoff $\Lambda$. For $I^0_{1f}$ one gets the regularized function
\begin{equation}
I^{vac\vphantom{g}}_{1f} = \dfrac{1}{2 \pi^2} \left[ \Lambda \sqrt{M_f^2 + \Lambda^2} +
M_f^2 \ln\left( \dfrac{M_f}{\Lambda+ \sqrt{M_f^2 + \Lambda^2}}\right) \right] \, . \label{I1freg}
\end{equation}
For $I^0_{2ff'}(q^2)$ we note that in order to determine the meson masses,
the external momenta $q$ in the loop integrals has to be extended to the region $q^2 < 0$.
Hence, we find it convenient to introduce $q^2=-q^2_m$, with $q_m > 0$. In this case the function has several poles. To treat them, we go from Euclidean to the original Minkowski space by taking $p_4=-i p_0$. Then, by choosing appropriate contours the $p_0$ integral can be calculated in the complex plane to yield
\begin{equation}
I^{vac\vphantom{g}}_{2ff'}(q^2) = - \dfrac{1}{8\pi^2 q_m^2} \int_0^\Lambda dp \: \dfrac{p^2}{p^2-r-i\epsilon}
\left[ \dfrac{q_m^2 + M_f^2 - M_{f'}^2}{\sqrt{p^2+M_f^2}} + \dfrac{q_m^2 - M_f^2 + M_{f'}^2}{\sqrt{p^2+M_{f'}^2}} \right] \, ,
\end{equation}
where
\begin{equation}
r= \dfrac{1}{4q_m^2} \left[ (M_f-M_{f'})^2 - q_m^2 \right] \left[ (M_f+M_{f'})^2 - q_m^2 \right] \, . \label{rrr}
\end{equation}
Depending on the value of $q_m$, this expression may still have a pole in a point of the integration line if $r>0$.
For those regions of $q_m$ where a pole exists, we proceed by employing a generalized version of the Sokhotski-Plemelj formula.
Assuming there exists a function $f(x)$ that has single poles at a set of values $x_j$, for which exist two other functions $g(x)$ and $h(x)$ such that $g(x_j)\neq 0$ and $h(x_j)\neq 0$, then
\begin{equation}
\lim_{\epsilon\rightarrow 0^+} \int_a^b dx \: \dfrac{h(x)}{f(x)+i\epsilon g(x)} = \mathrm{PV} \int_a^b dx \: \dfrac{h(x)}{f(x)}
- i\pi \sum_j \dfrac{h(x_j)}{|f'(x_j)|} \: \mathrm{sign} [g(x_j)] \, , \label{SPformula}
\end{equation}
where PV denotes the Cauchy principal value of the integral. By using this property we can fully calculate the complex function
$I^{vac\vphantom{g}}_{2ff'}$ in the most general case.
For the regularized real part we get
\begin{align}
Re\left[ I^{vac\vphantom{g}}_{2ff'}(-q_m^2)\right] = -\dfrac{1}{8\pi^2 q_m^2} \Bigg\{ &
(q_m^2 + M_f^2 - M_{f'}^2) \left[ \arcsinh \left(\dfrac{\Lambda}{M_f}\right) - F_f\right] + \nonumber \\
& (q_m^2 - M_f^2 + M_{f'}^2) \left[ \arcsinh \left(\dfrac{\Lambda}{M_{f'}}\right) - F_{f'} \right] \Bigg\} \, ,
\end{align}
where
\begin{align}
F_f =
\left\{
\begin{array}{ll}
  \dfrac{y_+}{\sqrt{M_f^2+y_+^2}}   \arctanh\left( \dfrac{\Lambda}{y_+} \sqrt{\dfrac{M_f^2+y_+^2}{M_f^2+\Lambda^2}} \right) & \mbox{for \ } q_m < q_m^{(0)} \mbox{ \  or \ } q_m > q_m^{(3)} \\
  \dfrac{y_+}{\sqrt{M_f^2+y_+^2}}   \arccoth\left( \dfrac{\Lambda}{y_+} \sqrt{\dfrac{M_f^2+y_+^2}{M_f^2+\Lambda^2}} \right) & \mbox{for \ } q_m^{(0)} <q_m < q_m^{(1)} \mbox{ \  or \ } q_m^{(2)} < q_m < q_m^{(3)} \\
  \dfrac{y_-}{\sqrt{M_f^2-y_-^2}}   \arctan\left( \dfrac{\Lambda}{y_-} \sqrt{\dfrac{M_f^2-y_-^2}{M_f^2+\Lambda^2}} \right) & \mbox{for \ } q_m^{(1)} <q_m < q_m^{(2)}\\
\end{array}
\right. \, .
\end{align}
Here $y_\pm = \sqrt{\pm r}$, with $r$ defined in Eq.~\eqref{rrr}, and
\begin{equation}
q_m^{0 \choose 3} = \left[ M_f^2 + M_{f'}^2 + 2 \Lambda^2 \mp 2\sqrt{(\Lambda^2 + M_f^2)(\Lambda^2 + M_{f'}^2)} \,\right]^{1/2}
\quad ; \quad q_m^{1 \choose 2} =  | M_f \mp M_{f'} | \, .
\end{equation}
For the regularized imaginary part we get
\begin{equation}
Im \left[ I^{vac\vphantom{g}}_{2ff'}(-q_m^2)\right] =
\left\{
\begin{array}{cl}
  -\dfrac{y_+}{4 \pi q_m} & \mbox{for \ } q_m^{(2)} <q_m < q_m^{(3)} \\
 0 & \mbox{otherwise}\\
\end{array}
\right. \, .
\end{equation}
Putting all together, the regularized version of the vacuum $c^0_{ff'}$ function defined in Eq.~\eqref{c0ff'} is given by
\begin{equation}
c^{vac\vphantom{g}}_{ff'} (q^2=-q_m^2) = 2 N_c \left\{ \dfrac{ I^{vac\vphantom{g}}_{1f} + I^{vac\vphantom{g}}_{1f'}}{2} -
\left[ q_m^2 - (M_f - M_{f'})^2 \right] I^{vac\vphantom{g}}_{2ff'}(-q_m^2) \right\}\, . \label{cff'reg}
\end{equation}

%%%%%%%%%%%%%%%%%%%%%%%%%%%%%%%%%%%%%%%%%%%%%%%

\section{Explicit expression of the neutral magnetic function $\mathbf{c^{mag}_{ff'}(q_\perp^2,q_\parallel^2)}$}

The unregularized neutral function $c_{ff'}$ in momentum space was originally defined in Eq.~\eqref{cff'mom}. Following a standard calculation (see~\cite{Coppola:2019uyr} for details) and assuming $Q_f=Q_{f'}$ we obtain
\begin{align}
c_{ff'}(q_\perp^2,q_\parallel^2) \: = \ & \dfrac{N_c B_f}{2\pi^2} \int_0^\infty dz \int_0^1 dy \:
\exp\left\{-z\left[ y M_f^2 + (1-y) M_{f'}^2 + y(1-y) q_\parallel^2 -i\epsilon \right] \right\} \times \nonumber\\[2mm]
& \exp\left[ -\dfrac{q_\perp^2}{B_f} \gamma_f(y,z) \right] \bigg\{ \left[ M_f M_{f'} + \dfrac{1}{z} - y(1-y)q_\parallel^2 \right] \coth(z B_f)
\, + \nonumber\\[2mm]
& \dfrac{B_f}{\sinh^2(zB_f)} \left[ 1 - \dfrac{q_\perp^2}{B_f} \gamma_f(y,z) \right] \bigg\}  \, ,
\end{align}
where
\begin{equation}
\gamma_f(y,z) = \dfrac{\sinh(yzB_f)\sinh[(1-y)zB_f]}{\sinh(zB_f)} \, .
\end{equation}
As usual, here we have used the changes of variables $\tau = yz$ and $\tau' = (1-y)z$, $\tau$ and $\tau'$
being the integration parameters associated with the quark propagators as in Eq.~\eqref{sfp_schw}.
The $B\rightarrow 0$ limit of this expression $c_{ff'}^0$ is given by Eq.~\eqref{c0ff'PT}. Then, the finite magnetic contribution is defined within the MFIR scheme as the difference
\begin{equation}
c_{ff'}^{mag}(q_\perp^2,q_\parallel^2) \, \equiv \, c_{ff'}(q_\perp^2,q_\parallel^2) - c^0_{ff'}(q^2) \, .
\end{equation}
For the calculation of the meson masses we can take $q_\perp^2=0$ and $q_\parallel^2=-q_m^2$, with $q_m>0$. Assuming that $q_m<M_f+M_{f'}$, one can integrate by parts to write this function in the form
\begin{equation}
c_{ff'}^{mag} (q_\perp^2=0,q_\parallel^2=-q_m^2) = 2 N_c \left\{ \dfrac{ I^{mag}_{1f} + I^{mag}_{1f'}}{2} -
\left[ q_m^2 - (M_f - M_{f'})^2 \right] I^{mag}_{2ff'}(-q_m^2) \right\}\, ,
\end{equation}
where the integral $I^{mag}_{1f}$ is defined as
\begin{align}
I^{mag}_{1f}  &=  \dfrac{B_f}{4\pi^2} \int_0^\infty \dfrac{dz}{z} \:  e^{-2z\, x_f} \left( \coth z - \dfrac{1}{z} \right) \nonumber\\[2mm]
&=  \dfrac{B_f}{2\pi^2} \left[ \ln \Gamma(x_f) - \left(x_f - \dfrac{1}{2}\right) \ln x_f + x_f - \dfrac{\ln{2\pi}}{2} \right] \, , \label{I1magAppB}
\end{align}
where $x_f= M_f^2/(2B_f)$. On the other hand $I^{mag}_{2ff'}$ is given by
\begin{eqnarray}
I^{mag}_{2ff'}(-q_m^2) =  - \dfrac{1}{8\pi^2} \int_0^1 dy
\int_0^\infty dz  \:  e^{ -2z(\bar x_{ff'}-i\epsilon) }  \left(
\coth z - \dfrac{1}{z} \right) \label{i2uno}
\end{eqnarray}
with
\begin{equation}
x_{ff'} = \dfrac{yM_f^2 +(1-y)M_{f'}^2 -y(1-y)q_m^2}{2B_f} \, .
\label{xff'}
\end{equation}
When $q_m<M_f+M_{f'}$ we always have that $x_{ff'} > 0$. Then
function $I^{mag}_{2ff'}$ as given in Eq.(\ref{i2uno}) is
well-defined and can alternatively written as
\begin{eqnarray}
I^{mag}_{2ff'}(-q_m^2) = \dfrac{1}{8\pi^2} \, \int_{0}^{1} dy \left[ \psi(\bar x_{ff'}-i\epsilon) - \ln(\bar x_{ff'}-i\epsilon) + \dfrac{1}{2 (\bar x_{ff'}-i\epsilon)} \right]
\, ,
\label{ImagAppB}
\end{eqnarray}
where $\psi(x)$ is the digamma function. Note that in this case one can safely take the $\epsilon\rightarrow 0$ limit.

On the other hand, when $q_m>M_f+M_{f'}$ it happens that $\bar
x_{ff'}$ can be negative in the integration domain. In this case, the integral in Eq.~\eqref{i2uno} is not convergent.
However, one can still proceed by considering the analytic extension of the form given in Eq.~\eqref{ImagAppB}.
Since $\bar x_{ff'}$ is a positive quadratic function of $y$, it is immediate to see that
$\psi(\bar x_{ff'})$ has $N+1$ poles, where
\begin{equation}
N=\mathrm{Floor} \left\{ \dfrac{1}{2B_f} \left[1 - \left( \dfrac{M_f-M_{f'}}{q_m} \right)^2 \right]
                                         \left[\dfrac{q_m^2}{4} - \left( \dfrac{M_f+M_{f'}}{2} \right)^2 \right] \right\} \, .
\end{equation}
To proceed we first isolate the poles by using the digamma recurrence relation
\begin{equation}
\psi(\bar x_{ff'}-i\epsilon) = \psi(\bar x_{ff'}+N+1) - \sum_{n=0}^N \dfrac{1}{\bar x_{ff'}+n-i\epsilon} \, .
\end{equation}
Expressed this way, the first term in the right-hand side is pole-free. Then
\begin{align}
I^{mag}_{2ff'}(-q_m^2) = \dfrac{1}{8\pi^2} \, \int_{0}^{1} dy \left[ \psi(\bar x_{ff'}+N+1) - \ln(\bar x_{ff'}-i\epsilon) - \dfrac{1}{2}
\sum_{n=0}^N \dfrac{g_n}{\bar x_{ff'}+n-i\epsilon} \right] \, , \label{I2magBB}
\end{align}
where $g_n=2-\delta_{n0}$. The complex logarithm is defined by taking the principal branch. For the region where $\bar x_{ff'}<0$ we have
\begin{equation}
\lim_{\epsilon\rightarrow 0} \: \ln(-|\bar x_{ff'}|-i\epsilon) = \ln(|\bar x_{ff'}|) - i\pi \, .
\end{equation}
Lastly, the third term on the right-hand side of Eq.~\eqref{I2magBB} contains two simple poles, which once again can be handled using the generalization of the Sokhotski-Plemelj formula presented
in Eq.~\eqref{SPformula}. After some algebra we finally obtain that for $q_m>M_f+M_{f'}$
\begin{align}
I_{2ff'}^{mag}(q_\parallel^2=-q_m^2) =& - \dfrac{1}{8\pi^2} \Bigg\{
\ln \left[ \dfrac{\left( M_f \right)^{1-\alpha}\left( M_{f'} \right)^{1+\alpha}}{2 B_f} \right] + \dfrac{\beta_0}{2} \ln \left[\dfrac{\alpha^2 - ( 1+\beta_0)^2}{\alpha^2 - ( 1-\beta_0)^2}\right]
\nonumber\\[2mm]
& \hphantom{- \dfrac{1}{8\pi^2} \Bigg\{ } - 2 +
\dfrac{B_f }{q_m^2 }\sum_{n=0}^N \dfrac{ g_n }{\beta_n} \ln \left[\dfrac{\alpha^2 - ( 1-\beta_n)^2}{\alpha^2 - ( 1+\beta_n)^2}\right]\Bigg\} \nonumber\\[2mm]
& + \dfrac{1}{8\pi^2} \int_{0}^{1} dy  \ \psi(\bar x_{ff'} + N + 1) +
\dfrac{i}{8\pi} \left( \beta_0 -\dfrac{2 B_f }{q_m^2 } \sum_{n=0}^N \dfrac{g_n}{\beta_n} \right)\, , \label{I2BBfin}
\end{align}
with
\begin{equation}
\alpha = \dfrac{M_{f'}^2-M_f^2}{q_m^2}   \quad ; \quad \beta_n = \sqrt{ \left[ 1- \left(\dfrac{M_{f'}-M_f}{q_m}\right)^2 \right]\left[ 1- \left(\dfrac{M_{f'}+M_f}{q_m}\right)^2 \right] - \dfrac{8 n B_f}{q_m^2}}\, .
\end{equation}

We remark that the calculation of $I_{2ff'}^{mag}$ was performed here within the proper time formalism, which is well-defined for $q_m<M_f+M_{f'}$
and leads to Eq.(\ref{ImagAppB}). For $q_m>M_f+M_{f'}$ we have taken the analytic continuation of this equation.
As a consistency check, we have repeated the calculation using the Landau level representation for the quark propagator in Minkowski space, which is well-defined for all $q_m$, obtaining the same final result of Eq.~\eqref{I2BBfin}.

\bibliography{su3mesB_v4}

\end{document}